\documentclass{article}
\usepackage{spconf,graphicx}
\usepackage{amsmath,amssymb}
\usepackage{url,lipsum,cite}
\usepackage{multicol,multirow}
\usepackage{xpatch,xcolor}
\usepackage{tabularx}
\usepackage{makecell, booktabs}
\newcolumntype{Y}{>{\centering\arraybackslash}X}
\usepackage{hyperref}

\hypersetup{
    colorlinks=true,
    linkcolor=purple,
    filecolor=magenta,      
    urlcolor=purple,
}

\newcommand{\newpara}[1]{\vspace{4pt}\noindent\textbf{#1}}
\def\BigRoman{\uppercase\expandafter{\romannumeral\number\count 255 }}
\def\Romannumeral{\afterassignment\BigRoman\count255=}

\title{High-resolution embedding extractor for speaker diarisation}
\name{Hee-Soo Heo, Youngki Kwon, Bong-Jin Lee, You Jin Kim, and Jee-weon Jung}
\address{Naver Corporation, South Korea}
\begin{document}
\ninept
\maketitle
\begin{abstract}
Speaker embedding extractors significantly influence the performance of clustering-based speaker diarisation systems. 
Conventionally, only one embedding is extracted from each speech segment.
However, because of the sliding window approach, a segment easily includes two or more speakers owing to speaker change points. 
This study proposes a novel embedding extractor architecture, referred to as a high-resolution embedding extractor (HEE), which extracts multiple high-resolution embeddings from each speech segment. 
Hee consists of a feature-map extractor and an enhancer, where the enhancer with the self-attention mechanism is the key to success.
The enhancer of HEE replaces the aggregation process; instead of a global pooling layer, the enhancer combines relative information to each frame via attention leveraging the global context.
Extracted dense frame-level embeddings can each represent a speaker. Thus, multiple speakers can be represented by different frame-level features in each segment. 
We also propose an artificially generating mixture data training framework to train the proposed HEE. 
Through experiments on five evaluation sets, including four public datasets, the proposed HEE demonstrates at least $10\%$ improvement on each evaluation set, except for one dataset, which we analyse that rapid speaker changes less exist. 
\end{abstract}

\section{Introduction}
\label{sec:intro}
Speaker diarisation is the task of finding `who spoke when', which is an essential task for a range of applications, including speech dictation systems~\cite{anguera2012speaker,park2022review}.
The task is usually used as a pre-process of a speech recognition system to divide long speech recordings into short speaker-homogeneous segments~\cite{yu2017recognizing, kanda2019acoustic}.
The field is experiencing rapid breakthroughs accelerated by advances in deep learning, where diarisation error rate (DER) is widely adopted as the primary metric.

In recent literature, two flagship challenges exist in the speaker diarisation field: DIHARD~\cite{ryant2020third} and VoxSRC~\cite{nagrani2020voxsrc, brown2022voxsrc}. 
The systems reported in these challenges can be categorised into end-to-end deep neural network-based and clustering-based.
End-to-end neural diarisation (EEND) approach directly trains a speaker diarisation system from input speech~\cite{fujita2019end, fujita2019endasru}. 
This approach can make the system simple and easy to maintain.
However, in the current state, they generalise less than clustering-based models, despite their superior performances in some scenarios.
Most of the winning teams of recent challenges either adopt a clustering-based approach or a hybrid ensemble of both approaches~\cite{wangdku, cai2022kriston}. 

A typical clustering-based speaker diarisation system consists of three phases, each conducted by a sub-system: end point detection, speaker embedding extraction, and clustering.
An end point detection model detects short voice segments from the input audio (i.e., session), discarding non-speech regions.
Then a speaker embedding extractor extracts embeddings from the voice segments.
In particular, embeddings are extracted using a sliding window for each segment; for example, several studies extract embeddings with a $1.5$s window and $0.5$s shift.
Speaker labels are then assigned to these embeddings in the clustering phase. 

More reliable speaker embeddings can be derived from longer utterances with more information in speaker verification. 
However, in diarisation, this is not the case because the probability of an embedding including two or more speakers increases as the window size extends. 
These embeddings degrade diarisation performance because the clustering-based approach cannot assign multiple cluster labels to a single embedding.
A too-small window size is also harmful because speaker information is insufficient, resulting in malicious embeddings.

As addressed above, the window size impacts performance; hence, several studies have proposed methods accounting for this issue. 
Some studies utilise an automatic speech recognition model to detect speaker change points and adjust the window size accordingly while remaining speaker homogeneous segments~\cite{xia2022turn}. 
However, speech recognition models are often computationally burdensome since large models typically involve billions of parameters.
Others leverage multiple embeddings extracted from different window sizes~\cite{park2021multi, kwon2022multi, park2022multi}. 
Still, embeddings extracted near speaker change points can be less reliable. 

To overcome this problem, we propose a model referred to as high-resolution embedding extractor (HEE). 
The proposed model extracts speaker embeddings more frequently (e.g., $40$ embeddings with a $3.2$s window) without applying aggregation methods (e.g., attentive statistics pooling~\cite{okabe2018attentive}). 
Thus, HEE extracts multiple embeddings from each input segment, whereas a conventional embedding extractor extracts one embedding from each segment.
To account for the speaker change point issue, we further propose a training scheme for HEE using multi-speaker training samples. 
Hence HEE can deal with multi-speaker segments, and we can benefit from more extended window sizes while avoiding issues evoked by speaker change points.

The rest of this paper is organised as follows.
In Section~\ref{sec:Prop}, the overall speaker diarisation pipeline with architecture and training scheme of the proposed HEE is presented.
Experiments and corresponding results are discussed in Section \ref{sec:Exp} followed by a conclusion.

\section{Proposed system pipeline}
\label{sec:Prop}

Our speaker diarisation pipeline includes three sub-modules: HEE, feature enhancement module, and clustering module. 
Compared with the existing system, only the embedding extractor is changed in the proposed system~\cite{kwon2021adapting,kim2021disentangled}. 
First, we crop the input audio into voiced segments using reference end point information, in line with several preceding speaker diarisation studies~\cite{park2022multi, kwon2022multi, kwon2021adapting}. 
Second, HEE extracts speaker embeddings from each segment with a sliding window where the sizes of the window and shift are $3.2$ and $0.8$, respectively.
In this process, the overlapping embeddings extracted from each section are averaged. 
Third, a feature enhancement module, proposed in \cite{kwon2021adapting}, consists of an auto-encoder-based dimensionality reduction and attention-based aggregation that refines the embeddings.
This process accelerates the processing speed and improves performance by generating more compact representations and reducing noise from the affinity matrix composed of cosine similarities between embeddings. 
Last, spectral clustering algorithm~\cite{wang2018speaker, xia2022turn} assigns labels to each embedding, where the number of speakers is determined leveraging an eigenvalue threshold.

\subsection{HEE Architecture}
\label{ssec:Arch}

\begin{figure}[t!]
  \centering
  \includegraphics[width=0.95\columnwidth]{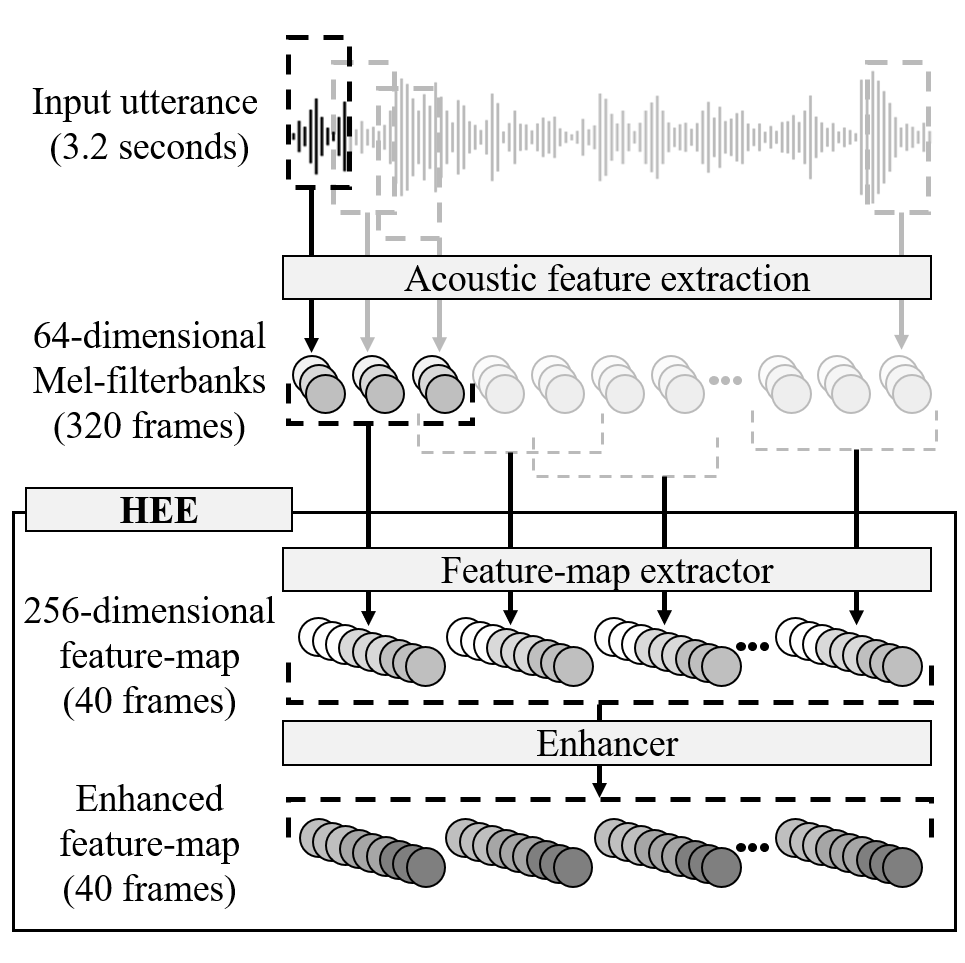}
  \caption{Architecture of the proposed high-resolution embedding extractor (HEE). The input mel-filterbanks are digested by a feature-map extractor, compressing eight frames into one frame. Then the enhancer leverages the global context based on the self-attention mechanism.}
  \label{fig:HEE_Arch}
  \vspace{-10pt}
\end{figure}

The architecture of HEE, illustrated in Figure~\ref{fig:HEE_Arch}, can be divided into two sub-modules: a feature-map extractor and an enhancer. 
The feature-map extractor represents speaker information in the input signal, similar to conventional speaker embedding extractors. 
However, we remove the aggregation process, typically conducted with global pooling layers.
Thus, multiple embeddings are extracted in proportion to the configured duration of the input length.
In particular, we adopt a window size of $3.2$ seconds which is more than double the widely adopted $1.5$ seconds configuration; embeddings are extracted every $80$ milliseconds resulting in $40$ embeddings. 

The enhancer inputs the feature-map extractor's output.
Its output has the same size as the input.
Two kinds of operations are expected from the enhancer, conditional to the input feature-map.
If one speaker exists in the input, features will be enhanced using the global information which spans the input. 
In contrast, when two or more speakers are included in the input, the enhancer identifies each speaker's region, and features are enhanced so that different speakers' features can be distinguished.

In order to successfully represent each speaker in a segment with multiple speakers, it is necessary to aggregate information from corresponding regions, excluding other speakers.
The self-attention module of transformer-based architectures makes it possible to distinguish a specific area from the input sequence~\cite{vaswani2017attention}.
We use the conformer, one of the transformer variations successfully applied in speaker verification~\cite{zhang2022mfa, gulati2020conformer}.
The enhancer of HEE consists of five conformer encoder blocks and has a residual connection between the input and the output. 
Each block contains a pointwise convolution with an expansion factor of four.

\newpara{Differences and advantages.}
HEE has two main differences compared to a conventional speaker embedding extractor.
First, it outputs multiple embeddings, whereas conventional extractors extract a single embedding.
Second, an enhancer is additionally exploited, which digests feature-map extractor outputs. 
When using the proposed HEE, we expect two following advantages enabled by the enhancer. 
First, if only one speaker exists in the feature-map, it is possible to extract embeddings with a broader window (more than $1.5$ seconds) than conventional embedding extractors. 
It would improve both the credibility and discriminant power of extracted embeddings.
Second, when multiple speakers are in the input, each speaker can be represented adequately without being corrupted by speaker change points. 
These advantages make the enhancer inevitable; without the enhancer, each frame-level feature will not have enough information to represent speakers.
In other words, the enhancer delicately plays the role of the pooling layer, which has been removed from the existing embedding extractor. 

\begin{figure}[t!]
  \centering
  \includegraphics[width=0.95\columnwidth]{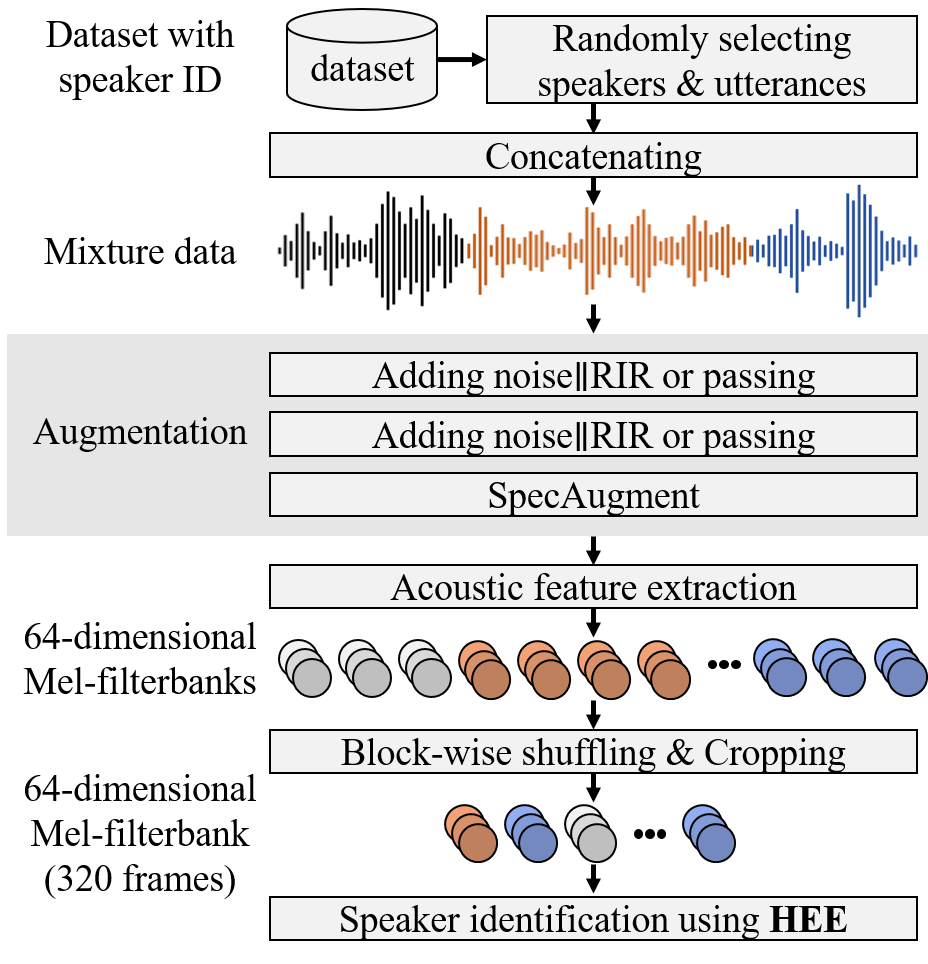}
  \caption{Data preparation process for HEE training. A mixture of data composed of a randomly selected number of speakers is concatenated. Data augmentation, mel-filterbank extraction, and proposed block-wise shuffling are applied in sequence.}
  \label{fig:HEE_framework}
  \vspace{-15pt}
\end{figure}

\subsection{Training framework}
\label{ssec:Train}
For HEE to operate as described in Section~\ref{ssec:Arch}, it is essential that HEE encounters abundant inputs where multiple speakers exist as well as single speaker inputs. 
We thus synthesise the training data using single speaker utterances for this condition.
Our training configuration is similar to that of EEND models~\cite{fujita2019endasru, horiguchi2020end}; the training of HEE can also be seen as a shorter (e.g., $3.2$s) local EEND training. 
However, there also exist a few differences.
First, our approach works on-the-fly, whereas EENDs use pre-generated synthetic data.
Second, we do not make overlaps, whereas EEND does. 
Third, our speaker labels are global, whereas EENDs adopt local speaker labels. 
Fourth, we apply clustering on the output of HEE, whereas EENDs directly use outputs as speaker labels. 

\begin{table}[!t]
  \centering
  \small
  \caption{
	Dataset statistics. We summarise the total duration in hours, the number of sessions and the average number of speakers per session.
	}
\begin{tabularx}{\columnwidth}{lYYY}
\Xhline{1pt}
\bf{Dataset} & \bf{Duration} & \bf{\# sessions} & \bf{\# speakers} \\
\hline
DIHARD {\Romannumeral 1}                    & 18.26         & 149       & 3.2   \\
DIHARD {\Romannumeral 2}                    & 22.42         & 194       & 3.3   \\
DIHARD {\Romannumeral 3}                    & 32.94         & 259       & 2.8   \\
AMI                         & 8.89          & 16        & 3.9   \\
In-house                    & 8.28          & 10        & 8.3  \\
\Xhline{1pt}
\end{tabularx}
  \label{tab:db}
\end{table}

Below, we detail the data preparation process of HEE training. 
First, the number of speakers composing the data is randomly selected between one to four for each training sample. 
Aiming for more challenging data, we set the probabilities of speakers being selected to $10\%$, $30\%$, $30\%$, and $30\%$ for one to four speakers, respectively.
Utterances from the selected speakers are concatenated to form speech of a specific length with speaker labels. 

However, the concatenated speech is not an informative training sample at this state because (i) each utterance was recorded in the different sessions and (ii) the mixture data contains a minimum number of speaker change points (e.g., two points for the three speaker case).
The model trained upon this data can leverage other information, especially channel differences, rather than being learned to represent speaker identities.

To counteract the first problem, we apply strong augmentation techniques to minimise the channel differences between connected audios. 
For strong augmentation, the process of adding noise or simulating room impulse response (RIR) with a certain probability is repeated twice.
We use reverberations and noises from simulated RIRs and MUSAN datasets~\cite{ko2017study,snyder2015musan}. 
In addition, SpecAugment~\cite{park2019specaugment} is also applied to mask a specific frequency band.
We use the augmentation recipe from the SpeechBrain library except for the range of the number of drops~\cite{ravanelli2021speechbrain}. 
We set the range from two to five for heavy augmentation. 

To mitigate the second problem, having a too small number of speaker changes, we shuffle the data in a block-wise manner, referred to as ``block-wise shuffling''. 
Here, block refers to a group of sequential frames which compose a duration between $0.5$ and $3$ seconds.\footnote{Note that labels are still assigned to each frame.}
Because directly editing raw waveforms can create discontinuities in the signal, 
we first extract acoustic features (e.g., mel-filterbanks) and perform block-wise shuffling to them.
We train HEE to perform speaker identification for each frame, where we remove the classification head after training is complete.
Figure~\ref{fig:HEE_framework} describes the data preparation process from which utterances are selected to where block-wise shuffling is conducted.

\newpara{Efficient implementation.}
We encountered a problem in training HEE where the mixture data could not be generated in real-time due to its complex process.
According to the aforementioned data preparation pipeline, numerous frames/blocks are discarded, which is inefficient. 
Multiple unused frames exist because the mixture data, including multiple speakers, inevitably exceeds $3.2$ seconds. 
Note that each audio of VoxCeleb, a widely used dataset with speaker ID, is at least three seconds~\cite{nagrani2017voxceleb}. 
We thus devise a more efficient practical data preparation pipeline.
First, a mixture data of $12.8$ seconds, four times longer duration, is composed, considering the existence of up to four speakers. 
Then, we generate four training samples for HEE.
This slight modification made data preparation much more efficient, and the training speed accelerated.

\section{Experiments}
\label{sec:Exp}

This section introduces the experimental settings and results. 
We adopt DER as the primary metric. 
We do not apply forgiveness collar for DER calculation, and overlaps are not ignored. 
We adopt reference end point information. 
Thus, miss (MS) in Table~\ref{tab:results} refers to the proportion of overlaps in the dataset.

\subsection{Datasets}
\label{ssec:dbs}

\newpara{Training datasets.} 
We use the development sets of VoxCeleb1\&2 datasets for training the proposed HEE~\cite{nagrani2017voxceleb,chung2018voxceleb2}. 
The training data is collected from YouTube videos and contains a wide range of domains.
These two development sets contain $1,211$ and $5,994$ speakers, respectively, resulting in $1,240,651$ utterances in total. 

\newpara{Evaluation datasets.} 
We use four public benchmark datasets and one in-house dataset to evaluate HEE. 
For the public datasets, we adopt test or evaluation sets of DIHARD {\Romannumeral 1}, {\Romannumeral 2}, {\Romannumeral 3} and AMI MixHeadset partition~\cite{mccowan2005ami, bredin2020pyannote}. 
We also evaluate an internally collected evaluation dataset.
This dataset does not include overlaps but is collected with rapid and abundant speaker change points. 
As a result, it achieves the worst speaker confusion (SC) among all five datasets in terms of baseline performance. 
Table~\ref{tab:db} summarises the statistics of each dataset. 
Based on this analysis, we predict that the in-house dataset containing the largest number of speakers will be the most challenging one.

\subsection{Configurations}
\label{ssec:exp_configs}
First, the feature-map extractor is pre-trained using the same recipe with~\cite{kwon2021adapting}, where the global pooling layer for aggregation is removed. 
After the pre-training is complete, the feature-map extractor is connected with a randomly initialised enhancer, and then trained again. 
The training consists of $20$ epochs, but the feature-map extractor is fine-tuned after the $10$'th epoch for stability. 
Each mini-batch contains $100$ samples, and each sample has $320$ frame-level mel-filterbank features. 
HEE is trained using Adam optimiser with an initial learning rate of $0.001$~\cite{kingma2015adam}. 
We use AAM-softmax~\cite{deng2019arcface} with a scaling factor of $30$ and a margin of $0.15$ as the loss function~\cite{chung2020defence}. 

\begin{table}[!t]
	\centering
	\small
	\caption{
	  Results on the DIHARD {\Romannumeral 1}, {\Romannumeral 2}, {\Romannumeral 3} evaluation sets, AMI test set, and in-house test set.  (FA: false alarm, MS: miss, SC: speaker confusion, \textbf{lower is better for all four metrics}).
	}
	\begin{tabularx}{\columnwidth}{lYYYY}
    \Xhline{1pt}
	 \textbf{Model} & \textbf{DER} & \textbf{FA} & \textbf{MS} & \textbf{SC}\\ 
    \Xhline{1pt}
	 \multicolumn{5} {c} {\bf{DIHARD {\Romannumeral 1}} }\\
     \hline\hline
	 Baseline (0.5s shift) & 19.41 & 0.0 & 8.71 & 10.70 \\
      Baseline (0.08s shift) & 19.23 & 0.0 & 8.71 & 10.52 \\
	 \hline
	 Challenge Winner~\cite{sell2018diarization} & 23.73 & - & - & - \\
	 Multi-scale GAT~\cite{kwon2022multi} & 19.00 & 0.0 & 8.71 & 10.29 \\
	 \hline
	 HEE (w/o enhancer)  & 45.71 & 0.0 & 8.71 & 36.91 \\
	 HEE &  18.30 & 0.0 & 8.71 & \textbf{9.59} \\
%	 \hline
%	 Track winner \cite{sell2018diarization} & 23.73 & - & - & - \\
    \Xhline{1pt}
    \multicolumn{5} {c} {\bf{DIHARD {\Romannumeral 2}} }\\
     \hline\hline
      Baseline (0.5s shift) & 19.97 & 0.0 & 9.69 & 10.28 \\
      Baseline (0.08s shift) & 20.04 & 0.0 & 9.69 & 10.35 \\
	 \hline
	 Multi-scale GAT~\cite{kwon2022multi} & 19.80 & 0.0 & 9.69 & \textbf{10.12} \\
	 Challenge Winner~\cite{landini2019but} & 18.42 & - & - & - \\
	 \hline
	 HEE (w/o enhancer)  & 45.83 & 0.0 & 9.69 & 36.11 \\
	 HEE &  19.98 & 0.0 & 9.69 & 10.29 \\
    \Xhline{1pt}
    \multicolumn{5} {c} {\bf{DIHARD {\Romannumeral 3}} }\\
    \hline\hline
      Baseline (0.5s shift) & 17.61 & 0.0 & 9.52 & 8.08 \\
      Baseline (0.08s shift) & 17.41 & 0.0 & 9.52 & 7.88 \\
	 \hline
	 Multi-scale GAT~\cite{kwon2022multi} & 17.35 & 0.0 & 9.52 & 7.83 \\
	 Challenge Winner~\cite{wang2021ustc} & 11.30 & - & - & - \\
	 \hline
	 HEE (w/o enhancer)  & 42.16 & 0.0 & 9.52 & 32.61 \\
	 HEE &  16.78 & 0.0 & 9.52 & \textbf{7.26} \\
    \Xhline{1pt}
    \multicolumn{5} {c} {\bf{AMI} }\\
    \hline\hline
      Baseline (0.5s shift) & 17.77 & 0.0 & 14.55 & 3.22 \\
      Baseline (0.08s shift) & 18.74 & 0.0 & 14.55 & 4.19 \\
	 \hline
	 MSDD~\cite{park2022multi} & 21.18 & - & - & - \\
      \hline
	 HEE (w/o enhancer)  & 53.08 & 0.0 & 14.55 & 38.51 \\
	 HEE &  17.30 & 0.0 & 14.55 & \textbf{2.75} \\
    \Xhline{1pt}
    \multicolumn{5} {c} {\bf{In-house test set} }\\
    \hline\hline
      Baseline (0.5s shift) & 20.46 & 0.0 & 0.0 & 20.46 \\
      Baseline (0.08s shift) & 14.23 & 0.0 & 0.0 & 14.23 \\
	 \hline
	 HEE (w/o enhancer)  & 68.40 & 0.0 & 0.0 & 68.40 \\
	 HEE &  9.95 & 0.0 & 0.0 & \textbf{9.95} \\
    \Xhline{1pt}
%    \multicolumn{5} {c} {\bf{VoxConverse} }\\
%    \hline
%      Baseline (0.5s sft) & 3.96 & 0.0 & 1.59 & 2.36 \\
%	 \hline
%	 HEE & 4.82 & 0.0 & 1.59 & 3.22 \\
%    \Xhline{1pt}
	\end{tabularx}
	\label{tab:results}
    \vspace{-14pt}
\end{table}

\subsection{Results}
\label{ssec:res}
Table~\ref{tab:results} addresses the main results in terms of DER. 
We analyse the results using speaker confusion, not the DER, because we experimented using reference end point detection information.
For the four datasets, except for the in-house dataset, we present three groups of systems. 
We provide two baselines, one or two recent state-of-the-art systems, and two proposed HEEs (without and with an enhancer).
Among two baselines, `Baseline ($0.5$s shift)' refers to the actual baseline. 
`Baseline ($0.08$s shift)' refers to modifying shift size to $0.08$ seconds, where it was built for in-depth analysis of HEE's effect; in other words, we wanted to verify whether HEE's performance difference is derived from changes in the shift size or not.

\newpara{Baseline vs Proposed.}
Four datasets out of five demonstrated performance improvement when using the proposed HEE; degradation on DIHARD {\Romannumeral 2}, the only dataset with no improvement, was neglectable ($10.28\%$ to $10.29\%$). 
Improvements on the three public datasets were consistent, where the average improvement was $10.26\%$.
The in-house dataset demonstrated surprisingly boosted the performance with $51.36\%$ improvement. 
In addition, we confirmed the impact of the enhancer throughout all datasets; HEE did not remain competitive without the enhancer.
 
In our observation, HEE was effective when altering the shift size in the baseline ($0.5$s to $0.08$s) was beneficial. 
For DIHARD {\Romannumeral 1} and {\Romannumeral 3}, AMI, and in-house test set, where decreasing the shift size was effective, HEE was also effective. 
Furthermore, in the in-house dataset where the improvement of different shift sizes was the most effective, HEE also demonstrated the most significant improvement; the in-house test set was indeed designed to test the rapid speaker change scenario. 
In contrast, for DIHARD {\Romannumeral 2}, where a decrease in shift size led to degradation, HEE also did not show improvement.
Densely extracting embeddings bringing improvement means more frequent/rapid speaker changes exist.
Hence, the proposed HEE's effectiveness is the most promising in challenging scenarios where rapid speaker changes often occur. 
In addition, a relatively $14.5\%$ performance improvement was confirmed even on the AMI dataset with low speaker confusion. 

\newpara{Comparison with state of the art.}
For all datasets, we compared the proposed HEE's performance with recent state-of-the-art systems~\cite{kwon2022multi, park2022multi,sell2018diarization,landini2019but,wang2021ustc}. 
HEE demonstrated state-of-the-art performance for DIHARD {\Romannumeral 1} and AMI.
However, for DIHARD {\Romannumeral 2} and DIHARD {\Romannumeral 3}, the winning entries which adopt ensembles outperformed HEE.

\newpara{Ablation results.}
We conducted two sets of ablation experiments on diverse augmentations and block-wise shuffling and reported the results in Table~\ref{tab:ablation}.
First, we differed the duration of HEE training and reported the results on the top four rows. 
The results showed that avoiding too short training duration was important; $2.4$s training degraded the performance $11\%$, compared to $3.2$s training.
In addition, configurations over a certain length always exhibited better performance than the baseline. 
Due to the limitations of GPU memory, we could not confirm lengths longer than $4.8$s.

Next, we verified the effect of each component of the HEE training framework by excluding one at a time. 
The results showed how much each component contributed to the performance improvement. 
Block-wise shuffling, which generates more speaker changes, had the most significant impact, followed by SpecAugment, which covers channel differences. 
All three components were essential for successful HEE training.

\begin{table}[!t]
\centering
	\small
	\caption{
	  Ablation experiments on the DIHARD {\Romannumeral 1} evaluation set. Different durations and components of HEE data preparation are explored. Results reported in SC.
	}
\begin{tabularx}{\columnwidth}{ccccY}
\Xhline{1pt}
\multicolumn{4}{c}{\bf{Configuations}}                      & \multirow{2}{*}{\bf{SC}} \\
\cmidrule{1-4}
\textbf{Duration} & \textbf{Noise$\parallel$RIR}  & \textbf{SpecAugment} & \textbf{Shuffling} &                     \\
\hline
2.4 s          & \checkmark             & \checkmark           & \checkmark        & 10.65                   \\
3.2 s          & \checkmark             & \checkmark           & \checkmark        & \bf{9.59}                   \\
4.0 s          & \checkmark             & \checkmark           & \checkmark        & 9.88                   \\
4.8 s          & \checkmark             & \checkmark           & \checkmark        & 9.72                   \\
\hline
3.2 s          &               & \checkmark           & \checkmark        & 10.05                   \\
3.2 s          & \checkmark             &             & \checkmark        & 10.92                   \\
3.2 s          & \checkmark             & \checkmark           &          & 11.53                  \\
\Xhline{1pt}
\end{tabularx}
  \label{tab:ablation}
  \vspace{-10pt}
\end{table}

\section{Conclusion and future works}
\label{sec:conclusion}

In clustering-based speaker diarisation, determining an embedding extractor's optimal input speech duration profoundly impacts performance but is difficult to select.
Too short inputs lessen embeddings' reliability, and too long inputs often become malicious owing to speaker changes. 
We proposed HEE, a system that can extract multiple embeddings from each segment by replacing the conventional global pooling layer with the proposed enhancer.
Owing to the self-attention mechanism, which can aggregate global context to each frame, the proposed HEE's outputs could remain in frame-level ($40$ frames for $3.2$s input) but were still discriminative.
Vast experiments on four public datasets and one in-house dataset confirmed that the proposed extractor could replace the existing embedding extractor and improve the performance. 
The improvement was even more significant for challenging conditions.
In the future, we plan to apply the proposed embedding extractor to an online speaker diarisation system.
We are also focusing on extending the HEE to cover the overlapped speech.

\clearpage
\bibliographystyle{IEEEbib}
\bibliography{shortstrings, refs}
\end{document}